\documentclass[12pt]{article}

\usepackage{graphics}
\usepackage{graphicx}
\begin{document}

\title{\Large
$N$-dependent Multiplicative-Noise Contributions 
in Finite $N$-unit Langevin Models: 
Augmented Moment Approach
\footnote{e-print: cond-mat/0512429}
}

\maketitle

\begin{center}
\author{\large Hideo Hasegawa
\footnote{e-mail: hasegawa@u-gakugei.ac.jp}}

{\it Department of Physics, Tokyo Gakugei University, \\
Koganei, Tokyo 184-8501, Japan}
\end{center}

\vspace{1cm}

\noindent
{\it \bf abstract}
{\small Finite $N$-unit Langevin models
with additive and multiplicative noises
have been studied with the use
of the augmented moment method (AMM)
previously proposed by the author
[H. Hasegawa, Phys. Rev E {\bf 67}, 041903 (2003)].
Original $N$-dimensional stochastic equations are
transformed to the three-dimensional deterministic
equations for means and fluctuations of
local and global variables.
Calculated results of our AMM are in good agreement
with those of direct simulations (DS).
We have shown that
although the effective strength of the additive noise of the
$N$-unit system is scaled as
$\beta(N)=\beta(1)/\sqrt{N}$, it is not the case
for multiplicative noise [$\alpha(N) \neq \alpha(1)/\sqrt{N}$],
where $\alpha(N)$ and $\beta(N)$ denote
the strength of multiplicative and additive noises,
respectively, for the size-$N$ system.
It has been pointed out that the naive assumption of 
$\alpha(N) = \alpha(1)/\sqrt{N}$
leads to result which violates the central-limit theorem 
and which does not agree with those of DS and AMM.}

\newpage

It has been recognized that 
stochastic (Langevin) equations subjected 
to additive and/or multiplicative noises
may be good models for discussing the property
of many systems not only in physics but also in
biology, chemistry, economy and networks.
In recent years, much attention has been paid
to multiplicative noise whose interesting aspects
have been intensively investigated 
(for a recent review, see ref. 1: related references therein).
It has been shown that multiplicative noise 
induces the phase transition where
multiplicative noise creates an ordered state
while additive noise generally 
works to destroy the ordered state.
The effect of multiplicative noise in Langevin equation
has been discussed in connection with the non-extensive
thermodynamics \cite{Tsallis88,Tsallis98}: non-Gaussian
distribution is realized for Gaussian multiplicative noise
\cite{Sakaguchi01,Anten02,Hasegawa05b}.

The property of multiplicative noise is different from that
of additive noise in some respects.
In order to make our discussion concrete, 
let's consider the simple, linear Langevin model
given by
\begin{equation}
\frac{dx_i(t)}{dt} = - \lambda \:x_i(t) + \alpha\: x_i(t)\: \eta_i(t)
+ \beta\: \xi_i(t),
\hspace{1cm}\mbox{($i=1-N$)}
\end{equation}
where $\lambda$ denotes the relaxation rate,
$\alpha$ and $\beta$ the strengths
of multiplicative and additive Gaussian white noises, respectively,
given by zero-mean random variables $\eta_i(t)$ and $\xi_i(t)$
with $< \eta_i(t) \:\eta_j(t')>
=< \xi_i(t) \:\xi_j(t')>=\delta_{ij}\delta(t-t')$.
The system given by eq. (1) has been studied by using the 
Fokker-Plank equation (FPE) mainly for $N=1$ or
$N=\infty$: for the latter case, the mean-field and
diffusion approximations are employed \cite{Munoz04,Risken96}.
In order to discuss the dynamics of the finite-$N$ Langevin model,
we define the global variable
given by $X(t)=(1/N)\:\sum_i x_i(t)$, and
sum eq. (1) over $i$ to get
\begin{equation}
\frac{dX(t)}{dt}=- \lambda \:X(t) + \alpha \:X(t) 
\left( \frac{\sum_i x_i(t) \:\eta_i(t)}{\sum_i x_i(t)} \right)
+ \frac{\beta}{N} \sum_i \xi_i(t).
\end{equation}
It is reasonable to replace the additive noise (third) term in eq. (2) by
$\beta(N) \xi$ where $\xi$ denotes the white noise with
\begin{eqnarray}
\beta(N)^2 &\equiv& 
\left< \left( \frac{\beta}{N} \sum_i \xi_i \right)^2 \right>
= \frac{\beta^2}{N}.
\end{eqnarray}
As for the multiplicative noise (second) term in eq. (2),
Mu$\tilde{{\rm n}}$oz, Colaiori and Castellano claimed in a recent paper 
\cite{Munoz05}
that it may be replaced by
$\alpha(N) X \eta$ with the white noise $\eta$
in the weak-noise case with
\begin{equation}
\alpha(N)^2 \equiv 
\left< \left( \alpha \frac{\sum_i x_i \:\eta_i}{\sum_i x_i} \right)^2 \right>
= \frac{\alpha^2}{N},
\end{equation}
if $x_i$ and $\eta_i$ are {\it uncorrelated}.
Equations (3) and (4) lead to the Langevin equation 
for $X$ given by
\begin{equation}
\frac{dX(t)}{dt}=- \lambda \:X(t) + \frac{\alpha}{\sqrt{N}} X(t) \: \eta(t)
+ \frac{\beta}{\sqrt{N}} \xi(t),
\end{equation}
from which the FPE for the finite-$N$ system is easily obtained.
The assumption given by eq. (4) naively seems reasonable.
It may be, however, not justified 
because $x_i$ and $\eta_i$ are {\it correlated}: 
$x_i$ may be increased (decreased) 
for $x_i > 0$ ($x_i < 0$) by a positive $\eta_i$ 
in eq. (1).

The purpose of this letter is to examine the
validity of the approximation given by eq. (4).
We will employ the augmented moment method (AMM)
(or dynamical mean-field approximation) 
developed by the author, 
which has been successfully
applied to a study of coupled stochastic systems 
\cite{Hasegawa03a}.
In this approach, we have discussed dynamics of coupled stochastic systems
in terms of a fairly small number of quantities relevant to
local and global variables, by transforming original stochastic equations to
deterministic equations. 
It is essentially the second-order moment method both for local 
and global variables. 
In ref. 9, 
we obtained equations of motions
for means, variances and covariances, by expanding variables
around their mean values. 
We have adopted, in this letter, an alternative FPE method
to obtain equations of motions of the quantities,
in order to avoid the difficulty
due to the Ito versus Stratonovich calculus inherent
for multiplicative noise. 
Numerical results calculated by using our AMM approach
are in good agreement with direct simulations (DS)
for the Langevin equation given by eq. (1).
On the contrary, the result obtained from the FPE using
the approximation given by eq. (4)
disagrees with those of DS and AMM.



We have adopted the finite $N$-unit general 
Langevin model given by
\begin{equation}
\frac{dx_i}{dt} = F(x_i) + \alpha \:G(x_i) \:\eta_i(t)
+ \beta \:\xi_i(t)+I_i^{(c)}(t),
\hspace{1cm}\mbox{($i=1-N$)}
\end{equation}
with 
\begin{equation}
I_i^{(c)}(t)=\frac{w}{Z} \sum_{k(\neq i)}[x_k(t)-x_i(t)],
\end{equation}
where $F(x)$ and $G(x)$ denote arbitrary functions of $x$,
$w$ the coupling strength, $Z=N-1$, 
$\alpha$ and $\beta$ denote the strengths of multiplicative
and additive noises, respectively, and
$\eta_i(t)$ and $\xi_i(t)$ express zero-mean Gaussian white
noises with correlations given by
\begin{eqnarray}
\langle \eta_i(t)\:\eta_j(t') \rangle
&=& \delta_{ij} \delta(t-t'),\\
\langle \xi_i(t)\:\xi_j(t') \rangle 
&=& \delta_{ij} \delta(t-t'),\\
\langle \eta_i(t)\:\xi_j(t') \rangle &=& 0.
\end{eqnarray}
The Fokker-Planck equation for the distribution of
$p(\{ x_i \},t)$ 
is given by 
\cite{Haken83}
\begin{eqnarray}
\frac{\partial}{\partial t}\: p(\{ x_i \},t)&=&
-\sum_k \frac{\partial}{\partial x_k}\{ [F(x_k) +
\frac{\phi\alpha^2}{2}G'(x_k)G(x_k) + I_k^{(c)}]\:p(\{ x_i \},t)\}  
\nonumber \\
&&+\frac{1}{2}\sum_{k }\frac{\partial^2}{\partial x_k^2} 
\{[\alpha^2 G(x_k)^2+\beta^2]\:p(\{ x_i \},t) \},
\end{eqnarray}
where $G'(x)=dG(x)/dx$, and
$\phi=1$ and 0 in the Stratonovich and Ito representations,
respectively.

When we consider the averaged, global variable $X$
given by
\begin{equation}
X(t)=\frac{1}{N} \sum_i x_i(t),
\end{equation}
the Fokker-Planck equation $P(X,t)$ for $X$ is given by
\begin{equation}
P(X,t) = \int \cdots \int \Pi_i \:dx_i \:p(\{x_i \},t)
\: \delta\Bigl(X-\frac{1}{N}\sum_i x_i\Bigr).
\end{equation}
%
The $k$-th moments of local and global variables are defined by 
\begin{eqnarray}
\langle x_i^k \rangle &=& 
\int \Pi_i \:dx_i \: p(\{x_i \},t) \:x_i^k, \\
\langle X^k \rangle&=& \int d\:X P(X) \:X^k. 
\hspace{1cm}\mbox{($k=1,2,\cdot \cdot$)}
\end{eqnarray}
By using eqs. (11) and (14), we get equations of motions
for mean, variance and covariance of local variable ($x_i$) 
given by
\begin{eqnarray}
\frac{d \langle x_i \rangle}{dt}
&=& \langle F(x_i) \rangle
+\frac{w}{Z}\sum_k [ \langle x_k \rangle - \langle  x_i \rangle]
+\frac{\phi \:\alpha^2}{2} \langle G'(x_i)G(x_i) \rangle, \\
%
\frac{d \langle x_i \:x_j \rangle}{dt}
&=& \langle x_i\:F(x_j) \rangle 
+ \langle x_j\: F(x_i) \rangle
+\frac{w}{Z}\sum_k [ \langle x_i x_k \rangle
+ \langle x_j x_k \rangle 
- \langle x_i^2 \rangle
- \langle x_j^2 \rangle] \nonumber \\
&+& \frac{\phi\:\alpha^2}{2}
[\langle x_i G'(x_j) G(x_j) \rangle
+ \langle x_j G'(x_i) G(x_i)\rangle] \nonumber \\ 
&+&[\alpha^2\:\langle G(x_i)^2 \rangle +\beta^2]\:\delta_{ij}. 
\end{eqnarray}
Equations of motions of mean and variance
of global variable ($X$) are obtainable 
by using eqs. (12), (16)-(17): 
\begin{eqnarray}
\frac{d \langle X \rangle}{dt}
&=&\frac{1}{N} \sum_i \frac{d \langle x_i \rangle}{dt}, \\
\frac{d \langle X^2 \rangle}{dt} 
&=& \frac{1}{N^2}\sum_i \sum_j 
\frac{d \langle x_i\:x_j \rangle}{dt}.
\end{eqnarray}
The mean-field approximation employs
only eq. (16) \cite{Broeck97}.
Equations (16) and (17) are adopted for a discussion
on the fluctuation-induced phase transition 
in infinite-$N$ stochastic systems \cite{Kawai04}.
Equations (18) and (19) play a crucial role
in discussing finite-$N$ systems, as will be shown below.
 
In the AMM \cite{Hasegawa03a},
we define the three quantities, $\mu$, $\gamma$ and $\rho$, given by
\begin{eqnarray}
\mu &=& <X> = \frac{1}{N} \sum_i <x_i>, \\
\gamma &=& \frac{1}{N} \sum_i <(x_i-\mu)^2>, \\
\rho &=& <(X-\mu)^2>.
\end{eqnarray}
It is noted that $\gamma$ expresses the averaged
fluctuations in local variables ($x_i$) 
while $\rho$ denotes fluctuations
in global variable ($X$).
Expanding $x_i$ in eqs. (16)-(19) around the average value 
of $\mu$ as
\begin{equation}
x_i=\mu+\delta x_i,
\end{equation}
and retaining up to the order of $<(\delta x_i)^2>$, we get
equations of motions for $\mu$, $\gamma$ and $\rho$ given by 
\begin{eqnarray}
\frac{d \mu}{dt}&=& f_0+f_2\gamma
+\left( \frac{\phi \: \alpha^2}{2}\right)
[g_0g_1+3(g_1g_2+g_0g_3)\gamma], \\
\frac{d \gamma}{dt} &=& 2f_1 \gamma
+ \left( \frac{2w N}{Z}\right) (\rho-\gamma) 
+(\phi+1) (g_1^2+2 g_0g_2)\alpha^2\gamma 
+ \alpha^2 g_0^2+\beta^2, \\
\frac{d \rho}{dt} &=& 2 f_1 \rho 
+(\phi+1)(g_1^2+2 g_0g_2)\alpha^2\rho
+ \frac{\alpha^2 g_0^2}{N}  + \frac{\beta^2}{N},
\end{eqnarray}
where $f_{\ell}=(1/\ell !)
\partial^{\ell} F(\mu)/\partial x^{\ell}$ and
$g_{\ell}=(1/\ell !) 
\partial^{\ell} G(\mu)/\partial x^{\ell}$.
Original $N$-dimensional stochastic equations given by eq. (6)
is transformed to three-dimensional deterministic equations
given by eqs. (24)-(26). 
In the limit of $\alpha=0$, they reduce 
to those obtained previously 
for the Langevin model subjected only to additive noises 
[eqs. (7)-(9) in ref. 9c].


It is easy to see that the stationary solutions
of $\gamma$ and $\rho$ for $w=0$
satisfy the central-limit theorem:
\begin{eqnarray}
\rho&=&\frac{\gamma}{N}, \\
\gamma &=& \frac{\alpha^2g_0^2+\beta^2}
{[-2f_1-(1+\phi)(g_1^2+2 g_0g_2) \alpha^2]}.
\end{eqnarray}
It is noted that even for $N = \infty$, 
local fluctuations exist ($\gamma \neq 0$),
although global fluctuations vanish ($\rho =0$).

Now we consider the result when the approximation
given by eq. (4) is employed (such result
is referred to as the APP hereafter).
The APP leads to
equations of motion for $\mu$ and $\rho$ given by
\begin{eqnarray}
\frac{d \mu}{dt}&=& f_0+f_2\gamma
+\left( \frac{\phi \: \alpha^2}{2N}\right)
[g_0g_1+3(g_1g_2+g_0g_3)\gamma], \\
\frac{d \rho}{dt} &=& 2 f_1 \rho 
+(1+\phi)(g_1^2+2 g_0g_2)\frac{\alpha^2 \rho}{N}
+ \frac{\alpha^2 g_0^2}{N}  + \frac{\beta^2}{N}.
\end{eqnarray}
We note that the third term of eq. (29)
and the second term of eq. (30) are different
from their counterparts in eqs. (24) and (26).
The ratio between $\rho$ and $\gamma$ in the stationary state
becomes
\begin{eqnarray}
\frac{\rho^{APP}}{\gamma^{APP}} &=& 
\frac{1}{N}
\left( \frac{-2 f_1 - (1+\phi)(g_1^2+2g_0 g_2)\alpha^2}
{-2 f_1-(1+\phi)(g_1^2+2g_0 g_2)\alpha^2/N} \right), \\
&\rightarrow& \frac{1}{N}
\left( \frac{-2 f_1 - (1+\phi)(g_1^2+2g_0 g_2)\alpha^2}
{-2 f_1} \right), 
\hspace{1cm}\mbox{(as $N \rightarrow \infty$)}
\end{eqnarray}
which is in contradiction to the central-limit theorem.


In order to make numerical calculations, 
we have adopted the linear Langevin model given by eq. (1): 
\begin{equation}
F(x)=-\lambda x,  \hspace{1cm}
G(x)= x. 
\end{equation}
In the Stratonovich representation, 
equations of motion given by eqs. (24)-(26) become
\begin{eqnarray}
\frac{d \mu}{dt}&=&-\lambda \mu 
+ \frac{\alpha^2 \mu}{2}, \\
\frac{d \gamma}{dt} &=& -2 \lambda \gamma 
+ \frac{2 w N}{Z}(\rho-\gamma) + 2 \alpha^2 \gamma
+ \alpha^2 \mu^2 + \beta^2, \\
\frac{d \rho}{dt} &=& -2\lambda \rho
+ 2 \alpha^2 \rho
+ \frac{\alpha^2 \mu^2}{N}  + \frac{\beta^2}{N}.
\end{eqnarray}
The stationary values for $w=0$ are given by
\begin{eqnarray}
\mu&=&0, \\
\gamma&=&\frac{\beta^2}{2(\lambda-\alpha^2)}, \\
\rho&=& \frac{\gamma}{N}.
\end{eqnarray}
Equation (38) shows that $ \alpha^2 < \lambda $ because $\gamma > 0$.

The probability $p(x)$ for the linear Langevin model ($N=1$)
in the stationary state is given by
\cite{Sakaguchi01,Anten02}
\begin{eqnarray}
p(x) &=& \frac{1}{Z} \left[ 1-(1-q)rx^2 \right]^{\frac{1}{1-q}},
\end{eqnarray}
with
\begin{eqnarray}
Z&=& \frac{1}{\sqrt{(q-1) r}} B\left(\frac{1}{2}, 
\frac{1}{q-1}-\frac{1}{2}\right), \\
r&=& \frac{2 \lambda+\alpha^2}{2 \beta^2}, \\
q&=& \frac{2 \lambda+3 \alpha^2}{2 \lambda +\alpha^2},
\end{eqnarray}
where $B(a,b)$ is the beta function.
By using $p(x)$ given by eqs. (40)-(43) with eq. (14), 
the first and second moments are given by 
\begin{eqnarray}
<x>&=&0, \\
<x^2>&=& \left[ \frac{1}{(q-1) r}\right]
\frac{B(\frac{3}{2},\frac{1}{q-1}-\frac{3}{2})}
{B(\frac{1}{2},\frac{1}{q-1}-\frac{1}{2})}, \\
&=& \frac{1}{r(5-3q)}=\frac{\beta^2}{2(\lambda-\alpha^2)}.
\end{eqnarray}
which agree with the result given by eqs. (37) and (38).

If we employ the APP, eqs. (30) yields
\begin{eqnarray}
\rho^{APP} &=& \frac{\beta^2/2N}{\lambda-\alpha^2/N}, \\
&\rightarrow& \frac{\beta^2}{2N \lambda}, 
\hspace{2cm}\mbox{(as $N \rightarrow \infty$)} \\
\frac{\rho^{APP}}{\gamma^{APP}} &=& 
\frac{1}{N}
\left( \frac{\lambda-\alpha^2}{\lambda -\alpha^2/N} \right), \\
&\rightarrow& \frac{1}{N}
\left( \frac{\lambda-\alpha^2}{\lambda} \right). 
\hspace{1cm}\mbox{(as $N \rightarrow \infty$)}
\end{eqnarray}
The value of $\rho^{APP}$ in eq. (48)
is independent of $\alpha$ for $N \rightarrow \infty$.
Equation (50) is in contradiction to the central-limit theorem.

In order to examine the validity of our AMM approach, we have made
direct simulations for the $N$-unit Langevin model given by eq. (1).
The Heun method is employed with
a time step of 0.0001.
Figure 1 shows the $N$ dependence of the stationary value of $\rho$ 
for the three sets of parameters: 
(1) $\alpha=0.8$, $\beta=1.0$,
(2) $\alpha=0.5$, $\beta=1.0$, and
(3) $\alpha=0.5$, $\beta=0.5$
with $\lambda=1$ and $w=0$.
Squares, circles and triangles show
results of direct simulations  (DS)
for the sets (1), (2) and (3), respectively. 
We note that $\rho$ follows the relation: $\rho \propto N^{-1}$.
In the case of $w=0$ under consideration, we get $\mu=0$ and
$\gamma(N)=\rho(N=1)$ independent of $N$.
Solid curves show the result of the AMM, which are in good agreement
with those of DS.
In contrast, the result with the APP
shown by dotted curves
disagrees with the result of DS and AMM.
For $N > 10$, $\rho^{APP}$ for the sets (1) and (2) 
becomes almost independent of $\alpha$ as eq. (48) shows.


We have  tried to obtain
the Fokker-Plank equation $P(X,t)$ for the global
variable $X$. We assume that the FPE for $P(X,t)$
is expressed by
\begin{eqnarray}
\frac{\partial P}{\partial t}
&=& - \frac{\partial}{\partial X}
\left\{ \left[F(X) + \frac{(\phi+A)\alpha^2}{2} 
G'(X) G(X) \right]\:P \right\} \nonumber\\ 
&&+\frac{\partial^2}{\partial X^2}
\left[\left(\frac{B \alpha^2}{2} G(X)^2
+ \frac{C \beta^2}{2}\right) P \right],
\end{eqnarray}
where $\phi=1$ and 0 in the Stratonovich
and Ito representations, respectively,
and $A$, $B$ and $C$ are parameters 
whose values will be determined
such as to yield the correct first and second moments,
as shown below. By using $P(X,t)$ in eq. (51), we get
the equation of motion for $\mu$ and $\rho$ given by
\begin{eqnarray}
\frac{d \mu}{dt}&=& f_0+f_2\gamma
+\left( \frac{(\phi+A) \: \alpha^2}{2}\right)
[g_0g_1+3(g_1g_2+g_0g_3)\gamma], \\
\frac{d \rho}{dt} &=& 2 f_1 \rho+ 
(\phi+A+B)(g_1^2+2 g_0g_2)\alpha^2\rho
+ B \alpha^2 g_0^2 + C \beta^2.
\end{eqnarray}
A comparison between eqs. (52) and (53) 
with eqs. (24) and (26) leads to
\begin{eqnarray}
\phi+A&=&\phi, \\
\phi+A+B&=&\phi+1, \\ 
B&=&C=\frac{1}{N}.
\end{eqnarray}
Unfortunately, conditions (54)-(56) are satisfied only for $N=1$
with $A=0$ and $B=C=1$,
but have no solutions for $N \neq 1$.
This implies that the FPE 
for the global variable $X$ 
in $N$-unit Langevin model is not 
expressed by the form given by eq. (51) 
with $A \propto N^{-\delta}$
for $N > 1$ ($\delta$: an index).
It is probable that $\sum_i x_i(t) \: \eta_i(t)$
in the second term of eq. (2)
may yield {\it non-Gaussian} multiplicative noise
because the distribution of $x_i(t)$
does not follow Gaussian.

To summarize, we have studied
finite $N$-unit Langevin model
including additive and multiplicative noises
by using AMM \cite{Hasegawa03a}, which has been reformulated 
with the use of FPE.
It has been pointed out that the scaling assumption given by
eq. (4) adopted in a recent paper (ref. 4) leads to results
violating the central-limit theorem [eq. 32] and 
disagreeing with those of DS and AMM.
The scaling relation of the effective strength against $N$
of multiplicative noise 
is different from that of additive noise.
Our AMM may be applied to the general Langevin model
with arbitrary forms of $F(x)$ and $G(x)$, and
also to multi-variable stochastic models like 
FitzHugh-Nagumo neuronal model.

A disadvantage of our AMM is that its applicability 
is limited to weak-noise cases. 
On the contrary, an advantage of the AMM is that we can easily discuss
dynamical property of the finite $N$-unit Langevin system
by solving the three-dimensional ordinary differential equations.
In contrast, within direct simulation and the FPE approach
we have to solve
the $N$-dimensional stochastic Langevin equations and
the $(N+1)$-dimensional partial differential 
equations, respectively.
Although the discussion presented in this letter is confined to the
stationary solution of equations of motions given by
eqs. (24)-(26) [or eqs. (34)-(36)],
it is possible to discuss the dynamical property of the
coupled Langevin model by solving them.
Actually in our previous papers 
\cite{Hasegawa03a},
we have studied the $N$-dependent
synchronization in networks described by
Langevin, FitzHugh-Nagumo and
Hodgkin-Huxley models subjected only to additive noises
with global, local or small-world couplings
(with and without transmission delays).
It is interesting to make such calculations by including
also multiplicative noises, which is left our future study.

\section*{Acknowledgements}
This work is partly supported by
a Grant-in-Aid for Scientific Research from the Japanese 
Ministry of Education, Culture, Sports, Science and Technology.



\newpage

\begin{figure}
\caption{
(Color online)
The $N$ dependence of the stationary $\rho$, fluctuations
in the global variable of $X$, for
the three sets of parameters:
(1) $\alpha=0.8$, $\beta=1.0$,
(2) $\alpha=0.5$, $\beta=1.0$, and
(3) $\alpha=0.5$, $\beta=0.5$
with $\lambda=1$ and $w=0$,
calculated by direct simulations with 1000 trials 
(DS: squares, circles and
triangles for (1), (2) and (3), respectively), 
the AMM (solid curves), and
the APP (dotted curves), bars denoting
the root-mean-square values of DS results.
}
\label{fig1}
\end{figure}

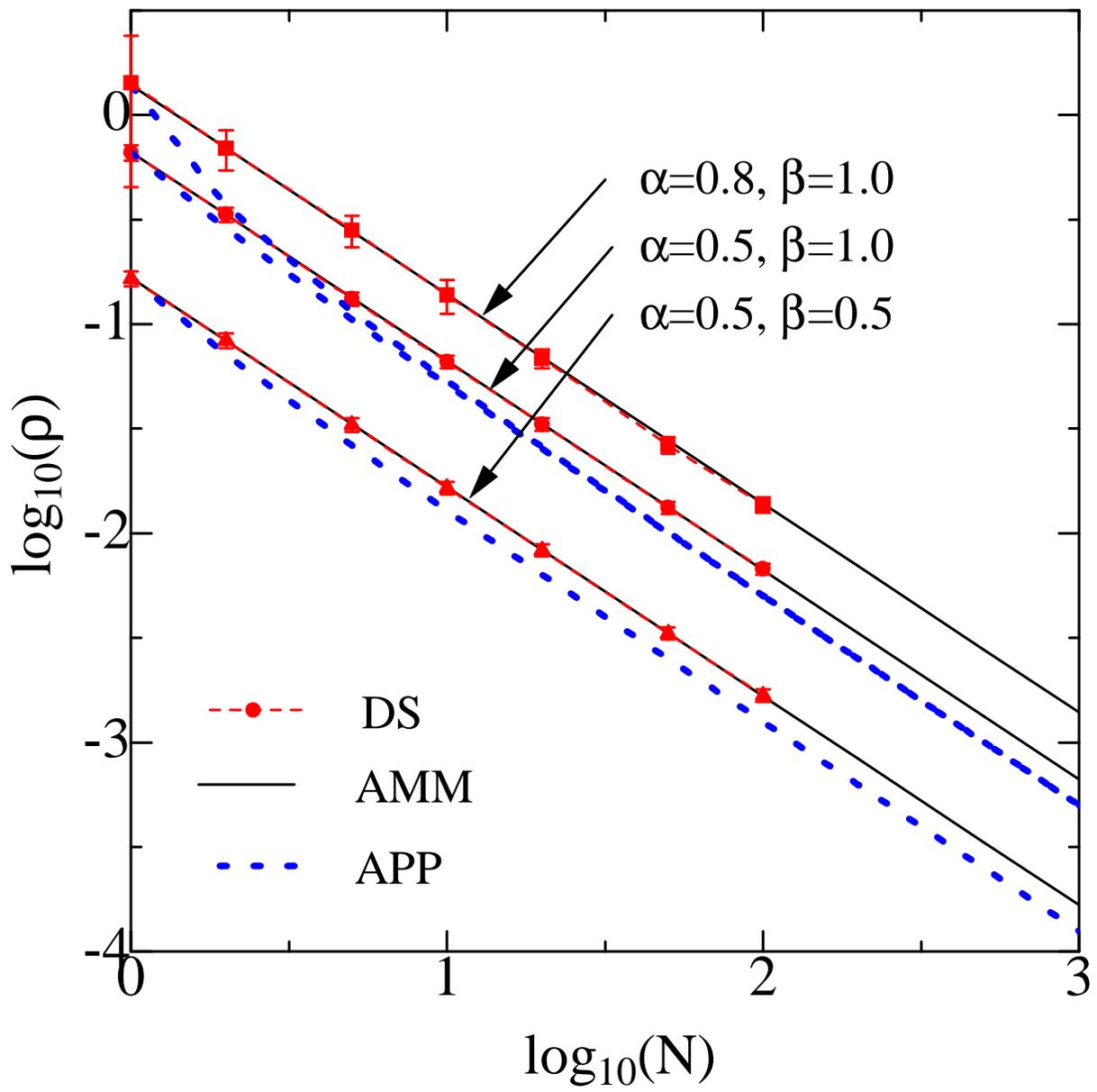
\end{document}